\documentclass[aps,pre, twocolumn, superscriptaddress, balancelastpage, showpacs, reprint,nofootinbib]{revtex4-1}

\usepackage{pifont}
\makeatletter
\newcommand{\fmarki}{*}
\newcommand{\fmarkii}{\ensuremath{\Davidstar}}
                
\def\@fnsymbol#1{{\ifcase#1\or \fmarki\or \fmarkii\or \fmarkiii\or \fmarkiv\or \fmarkv\or \fmarkvi\or \fmarkvii\or \fmarkviii\or \fmarkix \else\@ctrerr\fi}}
\makeatother

\renewcommand{\fmarki}{$\dagger$}
\renewcommand{\fmarkii}{$\ddagger$}

\usepackage[T1]{fontenc}
\usepackage{blindtext}
\usepackage{centernot}
\usepackage{graphicx}
\usepackage{amsmath,bbold}
\usepackage{times}
\usepackage{amssymb}
\usepackage{mathrsfs}
\usepackage{chemarr}
\usepackage{color}
\usepackage{url}
\usepackage{version}
\usepackage[hidelinks]{hyperref}
\usepackage{mwe,tikz}
\usepackage[percent]{overpic}
\usepackage{bm}
\usepackage[export]{adjustbox}
\definecolor{linkcolor}{rgb}{0,0,0.6} 
\definecolor{forestgreen}{rgb}{0.13, 0.55, 0.13}
\definecolor{frenchblue}{rgb}{0.0, 0.45, 0.73}
\definecolor{burntsienna}{rgb}{0.91, 0.45, 0.32}
\usepackage{algorithm}
\usepackage[noend]{algpseudocode}
\usepackage{booktabs}

\usepackage{array}
\usepackage{multirow}
\usepackage{subeqnarray}

\usepackage{enumitem}
\usepackage{cases}
\usepackage[normalem]{ulem}
\usepackage{transparent}

\newcommand\rsout{\bgroup\markoverwith{\textcolor{red}{\rule[0.5ex]{2pt}{0.8pt}}}\ULon}

\usepackage{lipsum}
\usetikzlibrary{patterns}

\newcolumntype{P}[1]{>{\centering\arraybackslash}p{#1}}
\newcolumntype{M}[1]{>{\centering\arraybackslash}m{#1}}

\begin{document}
  

\title{On the relative packing densities of pistachios and pistachio shells}

\author{Ruben Zakine}
\email{ruben.zakine@polytechnique.edu}
\affiliation{LadHyX, CNRS, École polytechnique, Institut Polytechnique de Paris, 91120 Palaiseau, France}
\affiliation{Chair of Econophysics and Complex Systems, \'Ecole polytechnique, 91128 Palaiseau Cedex, France}

\author{Michael Benzaquen}
\email{michael.benzaquen@polytechnique.edu}
\affiliation{LadHyX, CNRS, École polytechnique, Institut Polytechnique de Paris, 91120 Palaiseau, France}
\affiliation{Chair of Econophysics and Complex Systems, \'Ecole polytechnique, 91128 Palaiseau Cedex, France}
\affiliation{Capital Fund Management, 23 Rue de l’Universit\'e, 75007 Paris, France}

\date{\today}

\begin{abstract}
Given an appetizer bowl full of $N$ pistachios, what is the optimal size of the container  -- neither too small, nor too big -- needed for accommodating the resulting $2N$ non-edible pistachio shells? Performing a simple experiment we find that, provided the shells are densely packed, such container needs only be slightly more than half ($\approx 0.57$) that of the original pistachio bowl. If loosely packed this number increases to $\approx 0.73$. Our results are discussed in light of existing literature on packing ellipsoids and spherical shells.

\end{abstract}

\maketitle

\section{Introduction}

Humans have been eating pistachio seeds for over 8,000 years~\cite{marks2010encyclopedia,ireco}. Pistachios stem from \textit{Pistacia vera}, a small tree of the cashew family (\textit{Anacardiaceae})  believed to be indigenous to Persia (Iran)~\cite{braunencyclopedia}. Pistachios are even mentioned in the Old Testament: "So their father, Jacob, finally said to them, `If it must be, then do this: put some of the best products of the land in your bags (...) some pistachio nuts and almonds.'" (Genesis 43:11)~\cite{Bible2009}. 

While also used as an ingredient for sweets and other foods, pistachios are commonly served roasted and salted as pre-dinner snacks, in small appetizer bowls. A crucial issue for the host is that of finding the optimal container size needed to fit the non-edible pistachio shells. This is the question we address below.

Behind this seemingly quirky question stands a vast literature that tackles the issue of optimal packing of grains, colloids, and other particles, crucial to understand the emergent properties of various condensed matter systems~\cite{torquato2018}. As can be expected, the most studied objects are spheres, but even for this simple geometry, the century-long debate of their optimal packing is still topical today~\cite{hales_formal_2017}, as notably revealed by the 2022 Fields Medal laureate~\cite{viazovska_sphere_2017,cohn_sphere_2017}. The packing of non-rotationally invariant shapes is hardly mathematically tractable, which explains why it has only been explored recently by means of extensive numerical simulations in two or three dimensions~\cite{schilling_monte_2005,de_graaf_dense_2011,atkinson_maximally_2012, yuan_coupling_2018}, with comparisons to experiments in rare but noticeable situations, see e.g.~\cite{donev_improving_2004} for packing  ellipsoids. The intricate matter of packing polydisperse particles has also been explored, see e.g.~\cite{farr2009,yuan_coupling_2018}. 
In our case, the pistachios combine all the features that make their packing a particularly difficult one to study: pistachios are not uniform in size and shape, they are not rotationally invariant, and their shells are non-symmetric non-convex objects. For all these reasons, we decided to carry out an experimental study to answer the question raised above. A comparison to  existing literature on packing ellipsoids and spherical shells is provided in Section~\ref{sec:discussion}.

\section{Experimental analysis}
\label{sec:experiment}

We conduct an experiment allowing to measure the relative packing density of pistachios and pistachio shells, see pistachio geometry in Fig.~\ref{fig:schema}(a). In  Appendix~\ref{app:exp_data} we report some statistics computed over 50 randomly chosen pistachios  among the $N=613$  pistachios in our sample. Such statistics show that our sample is  weakly polydisperse with typical size dispersion of $\approx 10\%$.

Using a 2 litre graduated cylinder, we measure the loose ($n^\mathrm{loose}$) and dense ($n^\mathrm{dense}$) packing densities of our sample. The loose particle density is obtained by simply pouring the pistachios into the cylinder, thus leaving noticeable room in the packing, whereas the dense particle density is measured after shaking and rolling the cylinder for $\approx 20$ seconds (this ensured reproducible results). We perform the same measurement on the $2\times 613$ shells obtained after peeling all the pistachios and removing the seeds. Table~\ref{tab:fract} gathers the results. While there is little difference between the loose and dense packing densities for the full pistachios, that of the shells is $27\%$ larger after shaking. 
This can be easily understood by the fact that, contrary to the full pistachios, the shells can partially interlock when shaking the sample, see Fig.~\ref{fig:pistachio_shells}.

\begin{figure}[t!]
    \centering
    \includegraphics[width=\columnwidth]{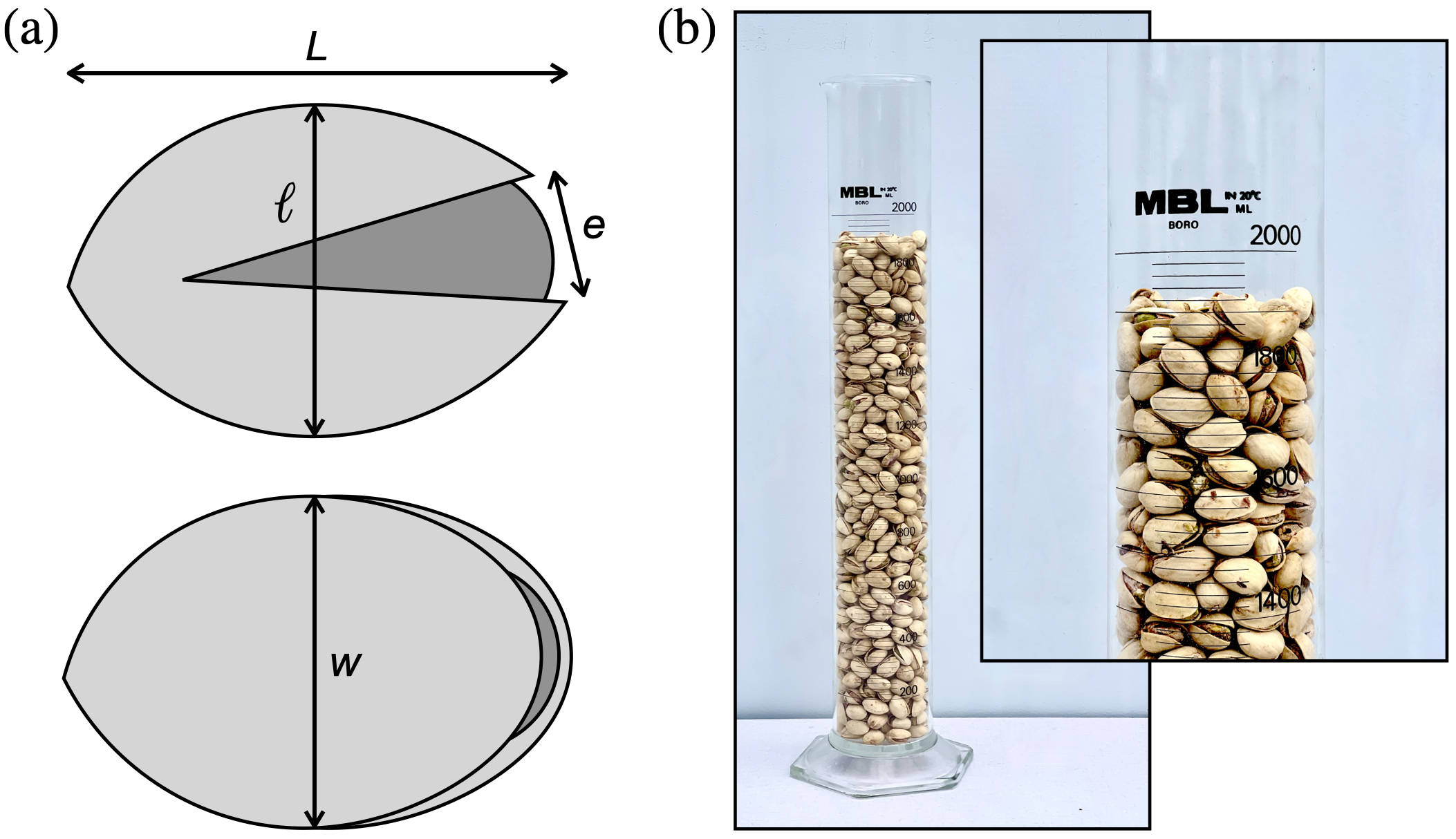}
    \caption{(a) Side and top views of a pistachio nut. Dimension statistics are given in Appendix~\ref{app:exp_data}.
    (b) Illustrative photograph of the experimental setup (613 pistachios in a 2 litre cylinder,  76\,mm diameter).}
    \label{fig:schema}
\end{figure}

\begin{table}[t]
\begin{tabular}{@{}lll@{}}
\toprule
 & loose  & dense \\ \midrule
                  $ n_{\mathrm{pist.}}$  & $313 \pm 6$ & $322 \pm 6$  \\
              $ n_{\mathrm{shells}} $  & $863 \pm 12$ & $1095\pm 12$ \\
              \bottomrule
\end{tabular}
  \caption{Loose and dense packing densities of full pistachios $ n_{\mathrm{pist.}}$ and shells only $ n_{\mathrm{shells}}$, in units of pistachios/litre. Error bars reflect reading uncertainty of volume in the cylindrical container.}
    \label{tab:fract}
\end{table}

To answer the question we initially raised, we assume that the full pistachios are served by simply pouring them into a bowl, consistent with the usual appetizer protocol, thereby achieving their loose packing density. For the shells, however, both loose and dense packing densities are relevant as the knowledgeable host may choose to shake the shell container as it fills to ensure maximum capacity. The relative size $\eta$ of the optimal shell container compared to the pistachio bowl is then simply given by the ratios of the particle densities:
\begin{eqnarray}
\eta^{\mathrm{loose/dense}} = \frac{2\times n_{\mathrm{pist.}}^{\mathrm{loose}}}{n_{\mathrm{shells}}^{\mathrm{loose/dense}}}.\label{eq:eta}
\end{eqnarray}
Injecting the numerical values of Table~\ref{tab:fract} into Eq.~\eqref{eq:eta}, one obtains:
\begin{subequations}
\begin{align}
&\eta^{\mathrm{loose}} =0.73 \pm 0.01  \label{eq:loosep}\\
&\eta^{\mathrm{dense}} =0.57 \pm 0.01 , \label{eq:densep}
\end{align}
\end{subequations}
meaning that, provided the shells are shaken, the size of the shell container needs only be slightly more than half that of the pistachio bowl.

Being concerned by the influence of the boundary conditions on our results we performed the same experiment with a wider cylinder (143 mm diameter), at the cost of lower precision. We obtained consistent results (within error bars).

\section{Discussion}
\label{sec:discussion}

Here we discuss our results in light of existing literature on the packing of non-spherical particles.

First, we confront the packing density of full pistachios estimated here to that of perfect ellipsoids of same dimensions $L$, $\ell$, and $w$. Using the parametrization $\alpha=L/w,\alpha^\beta=\ell/w$,  we find $\alpha = 1.5$ and $\beta = 0$,\footnote{One should bear in mind that pistachios are not true ellipsoids, and that there is a dispersion of pistachio sizes of $\approx 10\%$. For the sake of the argument we choose to work with the average values of $L$, $\ell$, and $w$. In full generality one has  $(1;\alpha^\beta=\ell/w;\alpha=L/w)=(1;1.00\pm0.05;1.5\pm0.1)$ where error bars reflect polydispersity.}  which  translates into a maximum  volume fraction of $\phi_\mathrm{ellipsoid}^{\max}=0.712$ for a random jammed packing of such ellipsoids, see Refs.~\cite{donev_underconstrained_2007, yuan_coupling_2018}. In our practical units, this is $n_\mathrm{pist.}^\mathrm{max}=364$ pistachios/litre, approximately $10\%$ higher than the packing density measured in our experiment. This is to be expected because, (i) while we do shake the cylinder to obtain a dense packing, the compaction procedure in~\cite{donev_underconstrained_2007} is  surely more effective,\footnote{In particular compaction algorithms most often discard static friction effects which in our system are expected to reduce mobility and thereby reduce the dense packing fraction. The same is to be expected for the compaction of shells.} and (ii) while not too significant, boundary effects due to the finite container diameter do exist and also go in the direction of decreasing the dense packing fraction. 

\begin{figure}
    \centering
\includegraphics[width=\columnwidth]{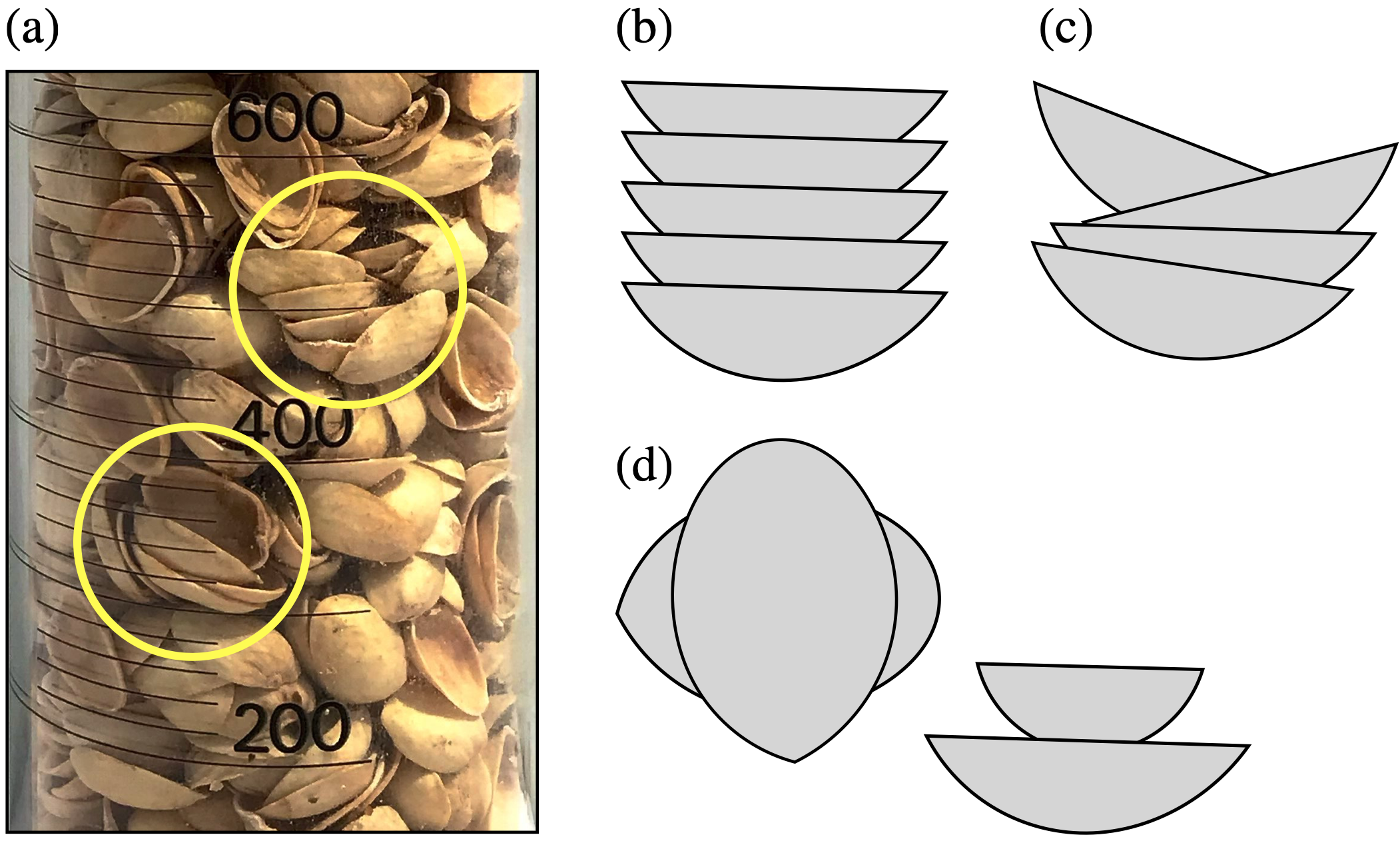}
    \caption{(a) Picture of densely packed pistachio shells in a 2 litre container. Yellow circles pinpoint two groups of interlocked pistachio shells (columnar phase). (b) and (c) Schematic view of interlocked pistachio shells stacked in one direction. (d) Top view of two pistachios unable to fully interlock because their main axes are not aligned.  }
    \label{fig:pistachio_shells}
\end{figure}

Second, discussing the results obtained for pistachio shells is more intricate. Indeed, while there are results for spherical shells~\cite{marechal_phase_2010}, there is, to our knowledge, no literature on their ellipsoidal counterparts. This being said, to push the analysis a bit further, we propose to define effective spherical shells which would have same volume and consistent steric hindrance.

The starting point is to assess the volume occupied by the pistachio shells, which we compute directly by pouring water in the graduated cylinder and measuring the volume difference with and without shells. The experiment yields a total volume of $360 \,\mathrm{mL}$ for $1226$ shells, hence a volume per shell of $v_\mathrm{shell} = 290\pm 5\, \mathrm{mm}^3$ where the error bar reflects reading uncertainty of volume. Following Ref.~\cite{marechal_phase_2010}, we define effective spherical shells with fixed diameter $\sigma = (L+w)/2$. Matching the volume of pistachio shells, we obtain an effective thickness $D_\mathrm{eff} = 1.37\,$mm (see Appendix~\ref{sec:sphericalcaps}).\footnote{Even though pistachio shells and spherical shells have the same volume by construction, $D_\mathrm{eff}$  is higher than the real pistachio shell thickness ($= 0.94 \pm 0.05$\,mm).
This is because, contrary to real shells,  effective spherical caps as defined in~\cite{marechal_phase_2010} have non-uniform radially decaying thickness; and what they call thickness actually corresponds the maximal thickness, see Fig.~\ref{fig:spherical}(a). Note that while it would be  interesting to explore the additional steric hindrance due to both irregular piling (Fig.~\ref{fig:pistachio_shells}(c)) and non-sphericity (Fig.~\ref{fig:pistachio_shells}(d)), this falls beyond the scope of the present analysis.} Computing $\sigma/D_\mathrm{eff}= 0.08$ and measuring (using water, see above) the volume fraction $\phi_\mathrm{shells}=0.327$ allows to  situate our system in the phase diagram displayed in Fig.~3 of Ref.~\cite{marechal_phase_2010}. Interestingly, we find that, in the language of Ref.~\cite{marechal_phase_2010}, our dense packing of pistachio shells lies in a regime of coexistence between the columnar phase (Fig.~\ref{fig:spherical}(b)) and the disordered (or \emph{fluid}) phase (Fig.~\ref{fig:spherical}(c)). This is remarkably consistent with our experiment findings: such coexistence can be clearly observed in Fig.~\ref{fig:pistachio_shells}(a). 

To conclude, our work contributes to the literature on the packing of non-spherical and non-convex objects. Future work should be devoted to assessing numerically our experimental results by simulating the packing of real pistachio shell geometries. 
We hope our results will help making the life of the appetizer host easier. See also Appendix~\ref{app:cost_effectiveness} for a short discussion on the cost effectiveness of shelled vs.~unshelled pistachios. Finally note that our analysis can be relevant in other situations, for instance to determine the optimal container needed to welcome mussel or oyster shells after a Pantagruelian seafood diner.

We deeply thank Alexander Rosenbaum for helping with counting and measuring pistachios during his visit to the Lab. We are grateful to Alexandre Darmon for fruitful discussions. This research was conducted within the Econophysics \& Complex Systems Research Chair, under the aegis of the Fondation du Risque, the Fondation de l’Ecole polytechnique, the Ecole polytechnique and Capital Fund Management.

\appendix

\section{Pistachio dimension statistics}
\label{app:exp_data}

\begin{figure}[h]
    \centering
    \includegraphics[width=\columnwidth]{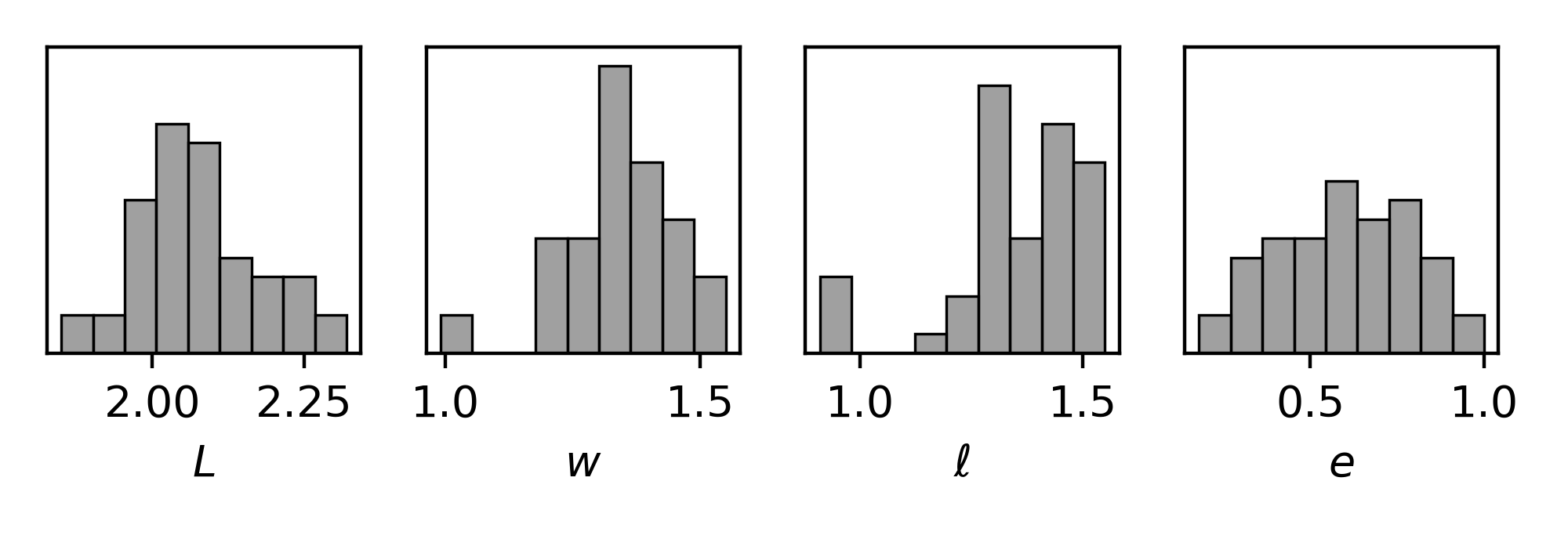}
    \caption{Pistachio dimension statistics (see Fig.~\ref{fig:schema}) computed over 50 randomly chosen pistachios in our sample.}
    \label{fig:stats}
\end{figure}

\begin{table}[h]
\begin{tabular}{@{}l||l@{}}
$ L $ \ & \ $20.74 \pm 1.04$ mm \\
$ w $ \  & \ $13.41 \pm 1.11$ mm  \\
$ \ell $ \ & \ $13.43 \pm 1.53$ mm \\
$ e $ \ & \ $5.99 \pm 1.89$ mm 
\end{tabular}
  \caption{Average pistachio dimensions ($\pm$ 1 standard deviation) computed over 50 randomly chosen pistachios  in our sample.}
    \label{tab:stats}
\end{table}

\section{Effective spherical shells}\label{sec:sphericalcaps}

\begin{figure}[h!]
    \centering
\includegraphics[width=\columnwidth]{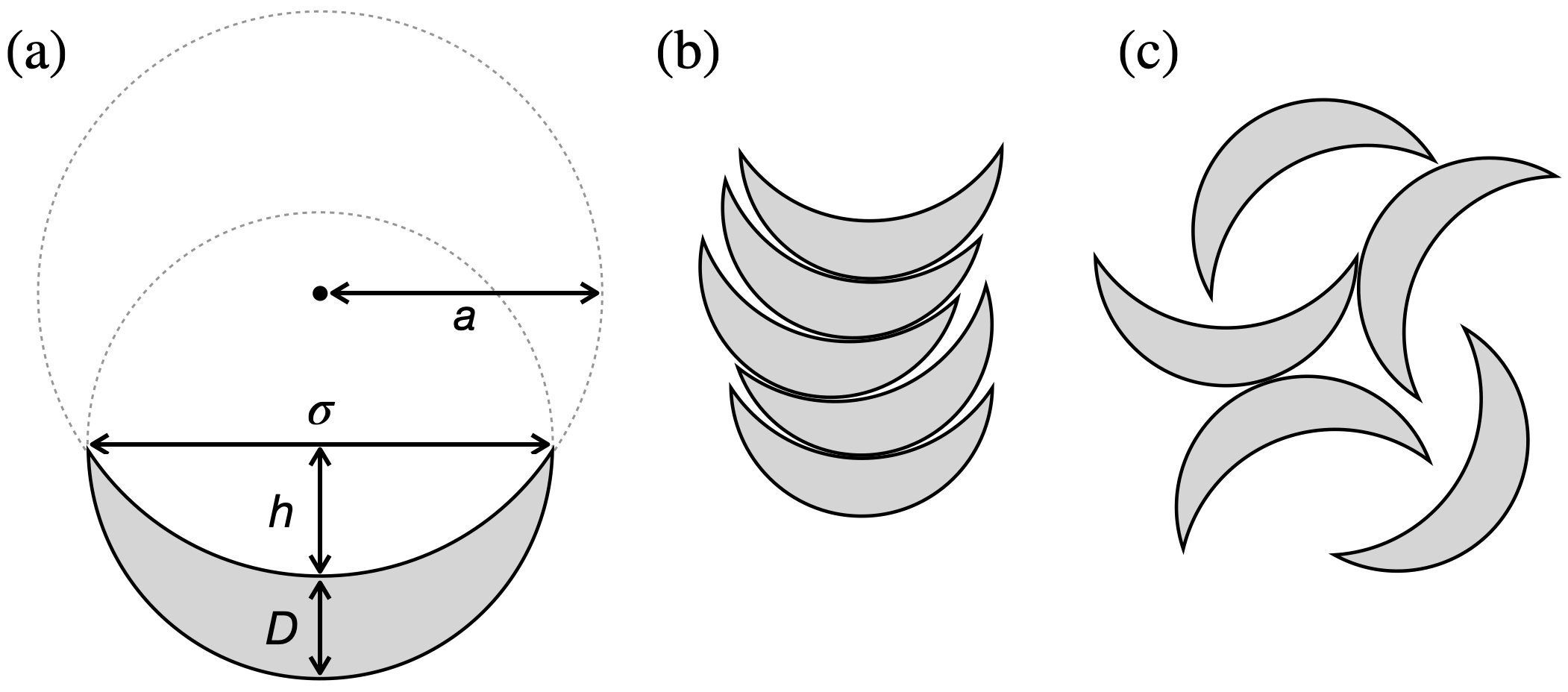}
    \caption{(a) Definition of spherical shells following Ref.~\cite{marechal_phase_2010}. (b) Sketch of the columnar phase made of piled up spherical shells. (c) Sketch of the disordered-fluid phase.}
    \label{fig:spherical}
\end{figure}

Here we present the shape of the spherical shells, as defined in \cite{marechal_phase_2010} (see Fig.~\ref{fig:spherical}), and compute their effective thickness such as to match the average volume of our pistachio shells.
The volume $v$ of a spherical shell (see Fig.~\ref{fig:spherical}(a)) writes
\begin{eqnarray}
    v = \frac23 \pi\left( \frac{\sigma}2\right)^2 - \pi h^2\left(a-\frac{h}{3}\right).
\end{eqnarray}
Using Pythagoras' theorem to eliminate $a$ and using $h=\sigma/2-D$ leads to 
\begin{eqnarray}
    v = \frac1{12} D\pi\left( 2D^2-3D\sigma +3\sigma^2\right).
\end{eqnarray}
Solving for $D$ using $v = v_\mathrm{shell}$ yields the value of the effective thickness (see above in the main text).

\section{Shelled or unshelled pistachios?}
\label{app:cost_effectiveness}

For the sake of completeness, let us also comment on the relative mass of the edible and non-edible parts of the pistachios. For $613$
pistachios, the mass of the shells is $381.1 \,\mathrm{g}$ while the edible mass is $438.9\, \mathrm{g}$; this means that only $54\%$ of a full dried pistachio is edible. 
This might come in handy when comparing the prices of full pistachios vs.~unshelled pistachios, as already argued by a few consumers online~\cite{blog_pistachio}, without neglecting in the cost-effective analysis the pleasure one might experience from actually pealing the pistachios. 

\clearpage

\bibliography{pistachios_biblio.bib}

\end{document}